\documentclass{article}

\usepackage[left=2.5cm, right=3cm, top=2.5cm]{geometry}
\usepackage{amsmath}
\usepackage{graphicx}
\usepackage{relsize}
\usepackage{epstopdf,epsfig}
\usepackage{afterpage}
\usepackage{amssymb}
\usepackage{color}
\usepackage{dcolumn}% Align table columns on decimal point
\usepackage{bm}% bold maate
\usepackage[square,numbers]{natbib}

\title{Reaction analogy based forcing for incompressible scalar turbulence}
\author{Don Daniel\thanks{dond@lanl.gov}, 
        Daniel Livescu\thanks{livescu@lanl.gov} \\
Los Alamos National Laboratory, USA\\
\and Jaiyoung Ryu\thanks{jairyu@cau.ac.kr} \\ Chung-Ang University, South Korea\\}

\date{}

\begin{document}

\newcommand{\uv}{\boldsymbol{u}}
\newcommand{\fv}{\boldsymbol{f}}
\newcommand{\xv}{\boldsymbol{x}}
\newcommand{\kv}{\boldsymbol{k}}
\newcommand{\Di}{\mathcal{D}}
\newcommand{\ufv}{\widehat{\boldsymbol{u}}}
\newcommand{\pf}{\widehat{p}}
\newcommand{\epst}{\epsilon_{target}}
\newcommand{\chit}{\chi_{target}}
\newcommand{\fra}{\mathrm{RA}}
\def\<{\langle}
\def\>{\rangle}

\maketitle

\begin{abstract}
We present a novel reaction analogy (RA) based forcing method for
generating stationary scalar fields in incompressible
turbulence. The new method can produce more general scalar PDFs (e.g.
double-delta) than current methods, while ensuring that scalar fields
remain bounded. Such features are useful for generating initial fields in non-premixed
combustion or for studying non-Gaussian scalar turbulence. The RA
method mathematically models hypothetical chemical reactions that
convert reactants in a mixed state back into its pure unmixed
components. Various types of chemical reactions are formulated and the
corresponding mathematical expressions derived such that the reaction term is smooth in the scalar space and is consistent with mass conservation. For large values of the
scalar dissipation rate, the method produces statistically steady
double-delta scalar PDFs.  Quasi-uniform, Gaussian, and stretched exponential scalar statistics are recovered for smaller values of the scalar dissipation rate. The shape of the scalar PDF can be further controlled by changing the stoichiometric coefficients of the reaction. The ability of the new method to produce fully developed passive scalar fields with quasi-Gaussian PDFs is also investigated, by exploring the convergence of the third order mixed structure function to the ``four-thirds" Yaglom's law. 
\end{abstract}

\section{Introduction}

Forced direct numerical simulations (DNS) provide detailed insights into the nature of homogeneous isotropic  turbulence 
\citep{alvelius1998,Lundgren2003, rubinstein_clark_livescu_luo_2004, Watanabe2004, Rosales_Meneveau05pf,Yeung2005,overholt_pope_1996,CSB2013}.
In contrast to free decay simulations, forced simulations
maximize the so-called `inertial' range of scales  on any given computational mesh. It can also be used to 
analyze flow responses when specific  scales of motion are excited. 
In this article, we propose a novel scalar forcing method based
on a chemical reaction analogy (RA) which is capable of 
producing more general passive scalar probability density functions (PDFs),
 for example double-delta, compared to classical forcing 
methods that are limited to producing Gaussian or near Gaussian scalar PDFs. 
The new method also ensures a vital property  of the scalar field: boundedness. 
 
In DNS studies of statistically stationary turbulence, various deterministic or random forcing terms usually mimic  the energy supplied to the system by the natural shear production mechanism in the turbulent kinetic energy equation, $\<{\bf u'} \cdot \nabla {\bf \overline{u}} \cdot {\bf u'}\>$, where ${\bf u'}$ are the velocity fluctuations and $\nabla {\bf \overline{u}}$ is the mean
velocity gradient. The velocity fluctuations can be replaced in the expression above with random noise or deterministic formulas, such that a specific form of the spectrum is obtained \cite{overholt_pope_1996}. When the actual velocity fluctuations are used in the forcing term and the mean velocity gradient is constant,  the forcing term becomes linear in velocity \cite{Lundgren2003,Rosales_Meneveau05pf,Petersen2010}. Typically, isotropic turbulence simulations are forced only at low wavenumbers. This maximizes the Reynolds number and ensures that the dynamics of the inertial range at higher wavenumbers evolve naturally, without being influenced by the details of the forcing~\cite{Kerr85jfm,Kaneda_Ishihara06jt}.

Scalar turbulence has been studied in the past by analogous linear forcing methods as those used for velocity turbulence, for example, by mimicking the production from the advection term corresponding to a constant mean scalar gradient \citep{overholt_pope_1996} or direct analogy with the linear velocity forcing term \citep{CSB2013}. The former method produces anisotropic scalar fields due to the directionality imposed by the mean scalar gradient, while the latter produces isotropic scalar fields. Neither method though, guarantees any bounds for the scalar field itself. While this may be relevant to some practical problems, there are many problems where the scalar quantities have 
natural bounds (e.g. mass or molar fractions vary between 0 and 1). The RA method proposed here, in contrast,  generates statistically steady scalar turbulence  by supplying energy to the scalar field via a nonlinear term analogous to a reaction rate. The method ensures that the scalar fields vary within
bounds specified at the onset. 

In addition, most studies to date of statistically stationary scalar turbulence concern Gaussian or quasi-Gaussian distributions, even though, generally, Gaussianity is not a general feature of scalar turbulence and it is known that scalar fields are more intermittent than the velocity fields \citep{Watanabe2004,kumar_girimaji_kerimo_2011}. Different types of scalar PDFs have been reported from laboratory experiments and numerical simulations of passive scalars which include stretched exponentials \citep{Jayesh1991, Jaberi1996, Shraiman2000}, pure Gaussian, and sub-Gaussian PDFs \citep{Warhaft2000,Ferchichi2002}. Many practical problems naturally exhibit non-Gaussian scalar distributions, as is the case with non-premixed combustion  \citep{Pope1979}.

The RA method uses a hypothetical chemical reaction to convert the mixed fluid back into unmixed pure states. Reactants are identified based on a fast reaction analogy similar to that proposed in Ref.
\cite{CD01} to quantify the width of the Rayleigh-Taylor mixing layer and further generalized and discussed in Ref. \cite{Livescu2008}. Here, we consider even more general reaction rates together with their physical constraints to be used as forcing terms, and discuss the resulting statistically stationary scalar fields and their dependence on the forcing term parameters. 

\begin{figure}
\begin{center}
\includegraphics[scale=0.14]{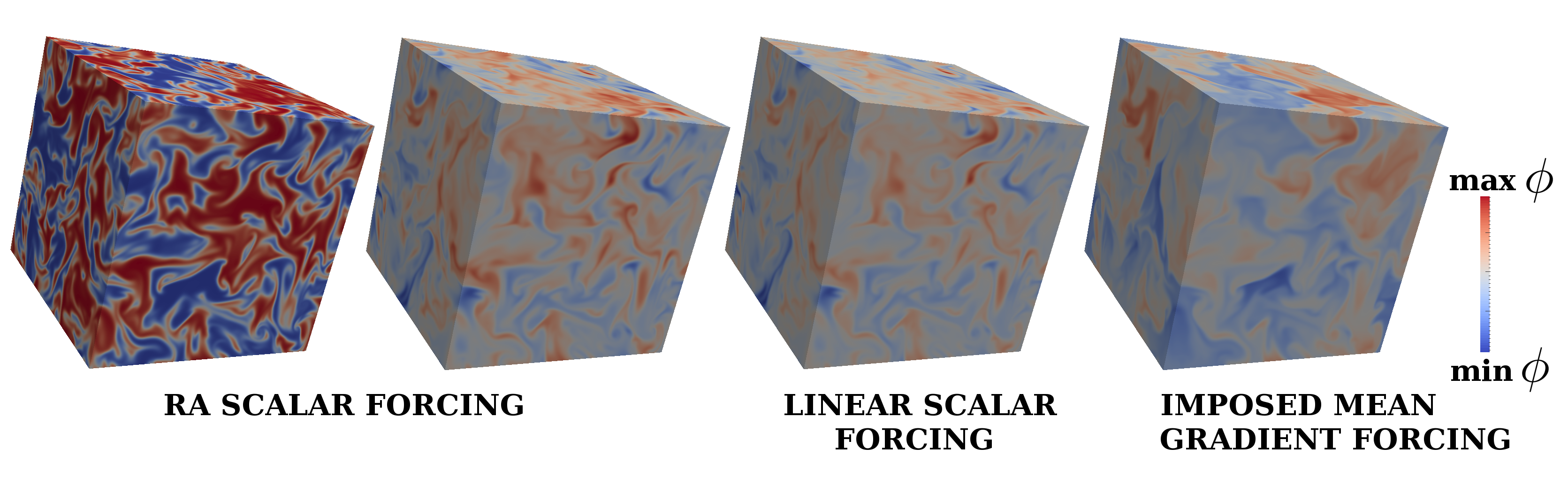}
\end{center}
\caption{Snapshots of the scalar field, $\phi$, produced by the proposed RA forcing (left two) and classical method of linear scalar forcing (third) and imposed mean gradient (fourth) methods. The underlying turbulent field is the same for all four cases.}
\label{fig:contours}
\end{figure}

To demonstrate the potential of the method, we illustrate the scalar fields produced by various scalar forcing methods in figure \ref{fig:contours}. 
 The RA method (left two) can produce both  
large concentrations of pure states (first figure, red and blue regions)
  or large concentrations of mixed states (second, white).
 In contrast,  classical methods of
linear scalar forcing (second from right) and imposed mean gradient 
forcing (right most) contain a large amount of 
mixed states.

A brief outline of the article is as follows.
The mathematical theory and formulations are discussed in \S 2. 
The results are described in \S 3 with conclusions given in \S 4.  
 
\section{Theory and Formulations}
We consider homogeneous isotropic turbulence governed by the incompressible constant density Navier-Stokes equations in a periodic box, subject to an explicit forcing term:

\begin{equation}
 \label{eq:org}
\nabla. \uv=0, 
 \quad  \frac{\partial \uv}{\partial t} + \uv. \nabla \uv =
 -\nabla p 
 +   \nu \Delta \uv   
 + \fv_{\uv}, 
\end{equation}  
where $\uv$ is the (zero mean) velocity vector, $p$ is the pressure normalized by the (constant) density and the forcing, $\fv_{\uv}$,  supplies energy at a constant input rate to a narrow band of wavenumbers, $\kv$, where $\| \kv \| < 1.5$. Maintaining a constant energy input rate has the advantage of being able to specify the Kolmogorov scale and, thus, the degree at which the flow is resolved, at the onset \citep{Petersen2010}.

The scalar field $\phi$ obeys an advection diffusion equation:
\begin{equation}
 \label{eq:scalar}
\frac{\partial{\phi}}{\partial t} + \uv . \nabla \phi = D \, \nabla^2  \phi  + \left[ \,\,  \underbrace{f(\uv)}_\mathrm{Mean \, Gradient} \mathrm{ or\,\,\, }   \underbrace{f(\phi)}_{\mathrm{linear \,or\, RA}} \,\, \right],
\end{equation}
where $f$ is the  scalar forcing term. In the equations above, the kinematic viscosity $\nu$ and diffusion coefficient $\Di$ are constant, with a Schimdt number of one, $Sc=\nu/\Di=1$. 

Previous scalar forcings  in Refs. \cite{overholt_pope_1996}  and \cite{CSB2013,PO2018} 
are linear and  consist of a  coefficient multiplied with either a velocity component, 
$u_i$, or the scalar field, $\phi$. The coefficient
for the former is a constant  mean scalar gradient
while for the latter is instantaneously calculated to 
maintain a constant scalar variance. For the RA method proposed here and explained below, $f$ is 
based solely on the scalar field. Note that mean scalar gradient forcing results in anisotropic scalar fields, while methods using $f(\phi)$ (including RA) produce isotropic scalar fields.

The RA method mathematically models fast chemical reactions occurring between
hypothetical reactants identified in a mixed fluid state.
Before identifying the reactants, let us define some useful metrics. 
 Consider a  mixed fluid state,
 in which the molar concentrations are such that there are
$N_A$ molecules of species $A$ and $N_B$ molecules
of species $B$ per unit volume.  
The molecular (molar)  masses of species 
$A$ and $B$ are  identical and the total number of molecules 
per unit volume is assumed to be a constant, $N=N_A+N_B$, which is required by incompressibility.
We define the molar fractions for species A as 
$X_A=N_A/N$ and for species B as $X_B=N_B/N$. 
The scalar field $\phi$ can then be related to the molar fractions of species A and B. If the scalar varies between zero and one, $\phi \in [0,1]$, then $X_A=\phi$ and $X_B=1-\phi$. When the scalar field varies from $\phi_l$ to $\phi_u$ such that $\phi \in [\phi_l,\phi_u]$, we have $X_A=(\phi-\phi_l)/(\phi_u-\phi_l)$ and $X_B=(\phi_u-\phi)/(\phi_u-\phi_l)$.
 
\begin{figure}
\begin{center}
(\emph{a})  \hspace{5cm}       (\emph{b}) \hspace{5cm}     (\emph{c})   \\
\includegraphics[width=5cm]{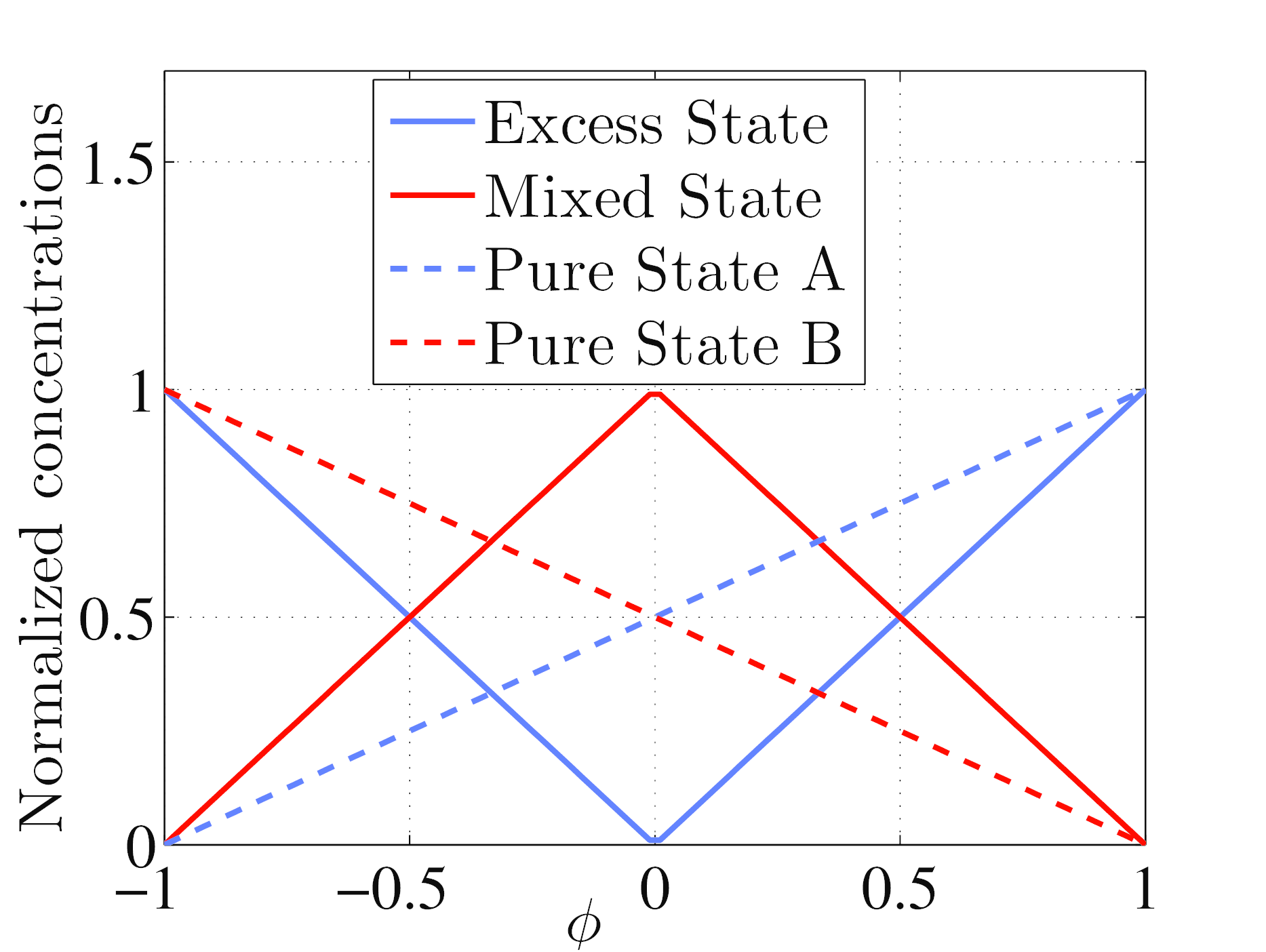}
\includegraphics[width=5cm]{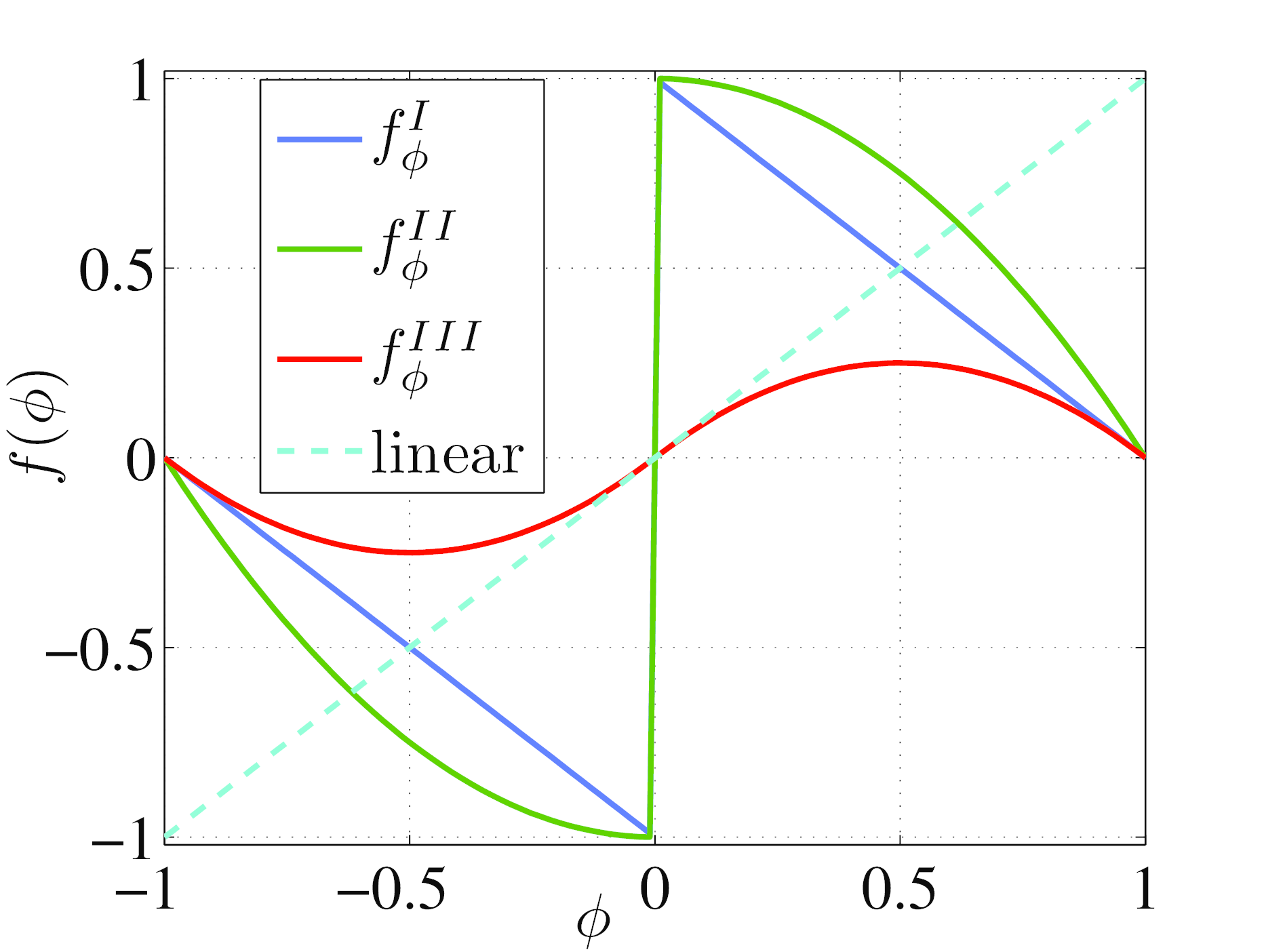}
\includegraphics[width=5cm]{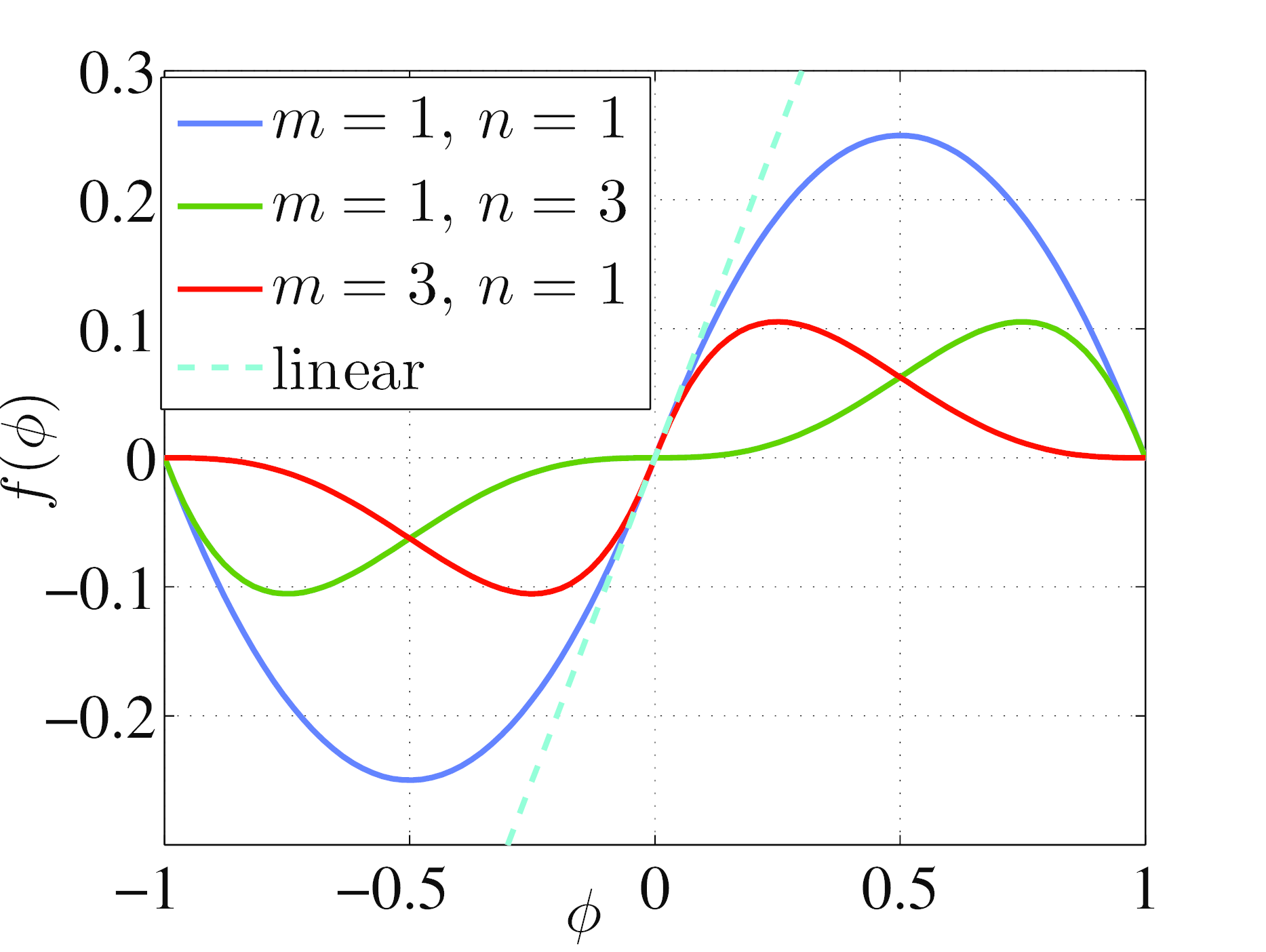}
\end{center}
\caption{(\emph{a}) Concentrations (normalized with maximum value) of
  mixed, excess and pure states versus $\phi$.
        (\emph{b}) Scalar reaction forcings of type A, type B and general
  forcing with $m=1$ and $n=1$ vs. $\phi$.
  (\emph{c}) General scalar reaction forcings,  $m=1$ and
  $n=1$, $m=1$ and $n=3$, and $m=3$ and $n=1$ vs. $\phi$.  }
\label{fig:forcings1}
\end{figure}

Various mathematical formulations can be derived to mimic
underlying reactions, see figure \ref{fig:forcings1}
for a plot of different formulations. 
A simple form in which reactants $A$ and $B$ get converted 
into each other is first discussed. 
When $N_A < N_B$, the preferred reaction is such that all of $A$ is 
 converted to $B$, increasing the concentration of $B$. 
When $N_A>N_B$, the  preferred reaction is such that all of $B$ 
is converted to $A$, increasing the concentration of $A$.  
This is represented by:
\begin{eqnarray}
A &\rightarrow& B    \quad \text{when $N_A < N_B$} \nonumber \\
B &\rightarrow& A   \quad  \text{when $N_A > N_B$}.         
\end{eqnarray}
The change in concentration of $A$ due to the first reaction
 when $X_A \leq X_B$ is 
$-K N  X_A$  and the second reaction when $X_A>X_B$ is 
$K N  X_B$; K is the reaction constant.
 Expressing in terms of scalar fluctuations,
 $\phi \in [-1,1]$ such that $\phi_l=-1$ and $\phi_u=1$, the forcing is:  
\begin{equation}
\label{eq:forcingA}
f_\phi^I= \begin{cases}
-f_c (1+\phi), \quad \, \text{ if $\phi < 0$}, \\
 f_c (1-\phi), \qquad   \text{if $\phi >0$},
\end{cases}
\end{equation}
where $f_c = K/2$ is the forcing coefficient, and the
forcing term has units of one over time.
The forcing term conserves the total number of molecules. Species $A$ and $B$ may 
physically represent two phases of the same 
chemical substance, for example the liquid
and vapor phases of H$_2$O, or just regions of high and low values or concentrations of the same scalar, for example temperature or a dye.

As a further example, consider a new series of chemical reactions, where 
$A$ reacts with
$B$ to produce more  $B$ when $X_A < X_B$, and $B$ reacts with
 $A$ to produce more $A$ when $X_A > X_B$:
\begin{eqnarray}
\label{eq:reactionB}
A + B &\rightarrow& 2 B    \quad \text{when $X_A < X_B$}, \nonumber \\      
A + B &\rightarrow& 2 A   \quad  \text{when $X_A > X_B$}.        
\end{eqnarray}
The change in concentration of $A$ is
$-K N^2 X_A X_B$  when $X_A \leq X_B$, and
$K N^2 X_A X_B$ when $X_A>X_B$.  The forcing in 
 terms of scalar fluctuations is:
\begin{equation}
\label{eq:forcingB}
f_\phi^{II}= \begin{cases}
-f_c(1-\phi)(1+\phi), \quad \, \text{ if $\phi < 0$}, \\
f_c(1-\phi)(1+\phi), \qquad   \text{if $\phi > 0$},
\end{cases}
\end{equation}
where $f_c=K/4$.
The forcing term is quadratic compared to the previous linear forcing for (3)-(4). 

Though conceptually simple to explain, we avoid using the above formulations because they are 
discontinuous at $\phi=0$ (figure \ref{fig:forcings1} a).  Consequently, we search for reactions and corresponding mathematical formulations  with better properties at $\phi=0$. To formulate reactions with smoother properties, we use ideas from a fast reaction analogy to describe the mixing from Refs. \cite{CD01,Livescu2008}. Consider hypothetical chemical species found within the mixing layer, the mixed state reactant,  $M$, and the excess state reactant, $E$. This can be either fluid $A$, i.e. $E_A$, or fluid $B$, i.e. $E_B$,  depending if the mixture is either $A$ or $B$ rich (figure \ref{fig:forcings1} b).

A simple formulation based on mixed and excess reactants is given by the following reactions:
\begin{eqnarray}
\label{eq:reactionC}
M + E_B \rightarrow 3 B    \quad \text{when $X_A < X_B$}, \nonumber \\
M + E_A \rightarrow 3 A   \quad  \text{when $X_A > X_B$}.      
\end{eqnarray}

The mixed state contains an equal number of $A$ and $B$ molecules, while the excess reactant can be either $A$ or $B$, such that the total equals the local concentrations of $A$ and $B$.
The goal of the reaction is to increase the
concentration of the excess pure states by decreasing 
the concentration of the mixed state. As a result, reactions
 (\ref{eq:reactionC}) constitute demixing processes that separate a
mixed state into its pure components.  

To find the forcing term, we  first express the reactant concentrations
in terms of mole fractions $X_A$ and $X_B$.
In regions where $X_A<X_B$, the mixed fluid will contain $N_A$ molecules of A and $N_A$ molecules of B per unit volume. Because one molecule of $M$ ($AB$)
is composed of one molecule each of $A$ and $B$, the 
 concentration of $M$ is given by  $[M]= N_A$. The remaining
molecules of $B$, i.e. $N-2 N_A$,  are present in the excess state, $E_B$, so that $[E_B]=N-2N_A=N_B-N_A$. 
Similarly when $X_A>X_B$, we have
$[M]= N_B$, $[E_A]= N_A-N_B$.
The time rate of increase of concentration of $E$ 
due to reaction between reactants $M$
and $E_{A,B}$ is
$2 K[M][E]$, i.e.,
$2 K N^2 X_A (X_A-X_B)$  when $X_A < X_B$ and
$2 K N^2  X_B (X_A-X_B) $  when $X_A>X_B$.  
Note that $d[E_A]/dt=d(N_A-N_B)/dt=d([A]-[B])/dt=2d[A]/dt$
and, similarly, $d[E_B]/dt=2 d[B]/dt$, which ensures the total number of molecules is conserved.
The forcing can be expressed in terms of scalar fluctuations as:
\begin{equation}
\label{eq:forcingC}
f_\phi^{III}= \begin{cases}
-f_c(-\phi)(1+\phi), \quad \, \text{ if $\phi \leq 0$}, \\
f_c \phi(1-\phi), \qquad   \text{if $\phi > 0$},
\end{cases}
\end{equation}
where $f_c=K/4$. The forcing term is smooth at $\phi=0$ (figure \ref{fig:forcings1} a).

A more generalized reaction model can also be considered, 
in which the mixed state reactant interacts with
 the excess state reactant in different stochiometric proportions:
\begin{eqnarray}
\label{eq:reactionG}
m M + n E_B &\rightarrow& (2m+n)B     
\,  \quad \text{when $X_A \leq X_B$} \nonumber \\  
m M + n E_A &\rightarrow&  (2m+n)A
   \quad  \text{when $X_A>X_B$},  
\end{eqnarray}
The corresponding forcing term is:
\begin{equation}
\label{eq:forcingG}
f_\phi^{IV} =  \begin{cases}
   - f_c (-\phi)^{n} (1+\phi)^m          \quad \text{when $\phi \leq 0$} \\
    f_c  \phi^n (1-\phi)^m           \quad \text{when $\phi>0$},  
\end{cases}
\end{equation}
where $f_c= m K /2^{m+n}$.
%The forcings are formulated such that the net molar
% concentration is conserved. In terms of $M$ and $E$,
%this is expressed  as $ f_{[M]}+ f_{[E]}/2=0 $.

Different choices of the  stochiometric coefficients, $m$ and $n$,
 result in unique  mathematical expressions. 
Choosing $M$ and $E$ as reactants lead to better smoothness properties compared to 
choosing $A$ and $B$.
  For example, when $m=3$ and $n=1$, (figure \ref{fig:forcings1} c),
 the forcing decreases to zero more quickly, with up to second derivatives vanishing
 at the scalar bounds.
Also, note that the forcings (\ref{eq:forcingA}), (\ref{eq:forcingB}),
(\ref{eq:forcingC}), (\ref{eq:forcingG}),  
reverse sign outside the scalar bounds, $\phi= \pm1$;
which ensures that the scalar field remains bounded. 

To eliminate the influence of small asymmetries in the
 initial scalar field, we further normalize the forcing expression.
If this normalization is not performed, a particular reaction is favored, 
transforming the entire scalar field 
to a trivial uniform state where either
$\phi=-1$ (or $\phi_l$ in the general case) or $\phi=1$ (or $\phi_u$ in the general case). To avoid this, 
the forcing is modified such that:
\begin{equation}
f_\phi^*=
\begin{cases}
f_\phi^{IV}/\|f_\phi^{IV}\|_{\phi \in [-1,0]} \quad
 \text{when $\phi \leq 0$} \\
f_\phi^{IV}/\|f_\phi^{IV}\|_{\phi \in (0,1]} \quad \text{when $\phi>0$.}
\end{cases}
\end{equation}
 This modification nullifies the effects of small scalar asymmetries in the initial 
configuration to keep the average zero, $\langle f_\phi^* \rangle=0$. Note $f_\phi=f_\phi^*$ in equation
(\ref{eq:scalar}).

The forcing coefficient $f_c$ is determined instantaneously to achieve a target scalar dissipation rate, $\chit$: 

\begin{equation}
\label{eq:targetchi}
f_c=\frac{\chit }
{ \langle  \phi f_\phi^* \rangle }. 
\end{equation}

The numerical results are obtained with a spectral version of the CFDNS code \cite{cfdns}, which follows a standard, fully dealiased pseudo-spectral algorithm using a combination of truncation and phase shifting, for incompressible Navier-Stokes equations with a passive scalar \cite{LJM00}. All simulations use a low band restriction ($\| \kv \| < 1.5$) of the linear velocity forcing with constant energy input \cite{Lundgren2003,Rosales_Meneveau05pf,Petersen2010}. The target velocity dissipation rate is calculated to satisfy a certain resolution condition. In the presence of a passive scalar, this condition yields:

\begin{equation}
\label{resolution}
\epst=
\begin{cases}
\frac{\Di^3}{(k_{max}\eta_{OC})^4_{target}} k_{max}^4 \qquad \text{ when $Sc \leq 1$} \\
\frac{\Di^2 \nu}{(k_{max}\eta_{B})^4_{target}} k_{max}^4 \qquad \text{ when $Sc > 1$ .} 
\end{cases}
\end{equation}
Here, $\eta_{OC,B}$ is the passive scalar micro-scale; the subscript OC refers to the Obukhov-Corrsin scale for $Sc < 1$, while subscript B refers to the Batchelor scale for $Sc > 1$. To demonstrate the method, all results presented correspond to $Sc=1$, which implies that $\eta$ and $\eta_{OC,B}$, the magnitudes of the  Kolmogorov and OC/Batchelor scales, are identical.
The target values of $(k_{max} \eta_{OC,B})_{target}$ and $ (k_{max} \eta)_{target}$
need to be large enough for the solution to be well resolved on a given grid. These are fixed at
either $1.5$ or $3.0$, as specified below. For simplicity, $\epst=1.0$ for all cases. To maintain the required resolution, viscosity and diffusion coefficients are changed appropriately to satisfy formula \ref{resolution}. Specifically, the simulations have been performed on $256^3$, $512^3$, and $1024^3$ meshes, resulting in Taylor Reynolds numbers ranging from $92$ to $410$. The turbulent kinetic energy at stationarity is approximately $3.0$. The results are averaged over $10$, $3$, and $1.5$ eddy turn over times for the three mesh levels, respectively. In terms of the time scale calculated based on the integral scale and kinetic energy, $\tau_L$, the results are averaged over at least $5 \tau_L$ intervals. 

$\chit$ is used to control the scalar PDF at stationarity. Most of the results below are obtained with the forcing following formula (\ref{eq:forcingG}), with stochiometric coefficients $m=1$ and $n=1$. This is essentially equivalent to formula (\ref{eq:forcingC}). The effects of different values for the stoichiometric coefficients are presented  in section (\ref{sec:mn}). To increase maximum allowable $\chit$, the scalar bounds used in the forcing terms are smaller (i.e. $\{-0.98,\ 0.98\}$) than the actual scalar bounds, $\{-1,\ 1\}$. This is explained in detail below.

\section{Results}

In this section, we present some sample results to highlight the properties of the new forcing method. 
First, the influence of the forcing parameters and resolution on the resulting scalar PDF are discussed. 
The spectral behavior of the forcing term is presented next, followed by a comparison with constant scalar gradient and linear scalar forcings in terms of scalar bounds, variance, and PDF. Finally, we present results showing the convergence to the 4/3 law for the mixed scalar-velocity third order structure function.

\subsection{The effects of forcing parameters on scalar PDF}

Here, we discuss the properties of the new RA forcing by examining how various forcing parameters  affect the stationary scalar probability density functions (PDFs), using $256^3$ simulations, with $\eta k_{max} =3.0$ ($\Delta x \sim \eta$), corresponding to $Re_{\lambda} \sim 92$. The main parameters controlling the properties of the forcing at stationarity are the target scalar dissipation rate and stoichiometric coefficients. This section also addresses the effect of resolution, using coarser  $256^3$ simulations with  $\eta k_{max} =1.5$ ($\Delta x \sim 2 \eta$), and $Re_{\lambda} \sim 154$, as well as $512^3$ and $1024^3$ simulations with $\eta k_{max} =3.0$ and $Re_{\lambda} \sim 143$ and $\sim 255$, respectively. Finally, we discuss the robustness of the method to changing the values of the scalar bounds. 

\subsubsection{Effects of varying the target scalar dissipation rate}       

\begin{figure}
\begin{center}
(\emph{a})  \hspace{5.9cm}  (\emph{b})  \\
\includegraphics[scale=0.3]{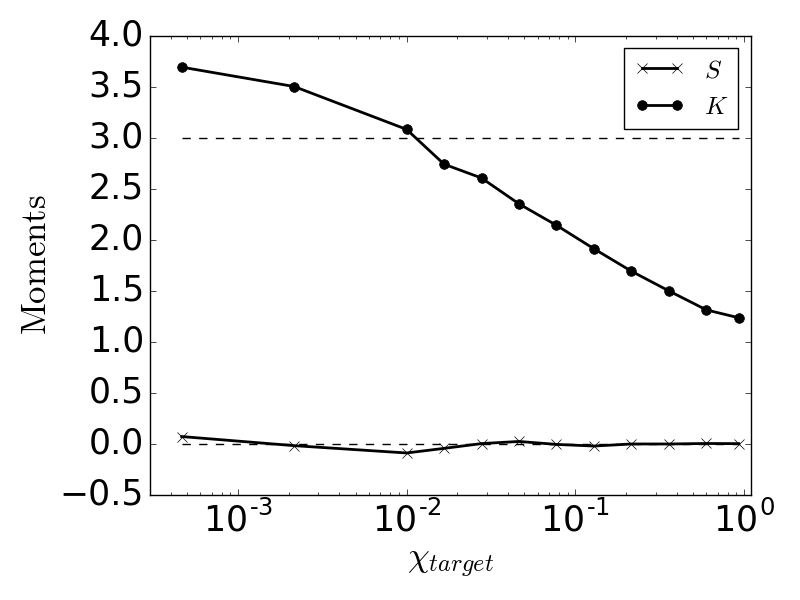}
\includegraphics[scale=0.3]{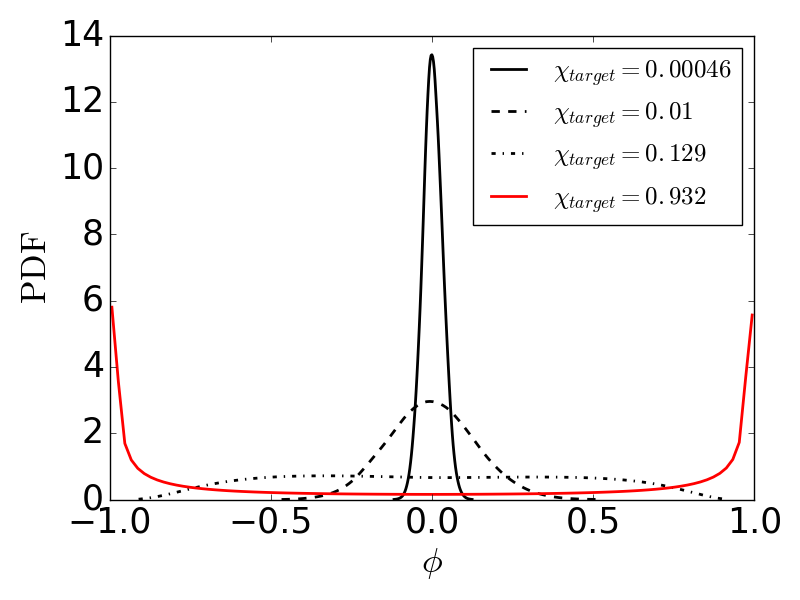}
\end{center}
\caption{(\emph{a}) Skewness S (crosses) and kurtosis (flatness) K (circles) scalar PDF moments obtained for various target dissipation rates, $\chit$ ranging from $0.00046$ to $0.932$. The Gaussian kurtosis of 3.0 is indicated with dashed lines. (\emph{b}) Stationary scalar PDFs.}
\label{fig:moments}
\end{figure}

Figure \ref{fig:moments}(\emph{a}) illustrates the skewness, $S=-\frac{\langle \phi^3 \rangle}{\langle \phi^2 \rangle^{3/2}}$, and kurtosis,  $K = \frac{\langle \phi^4 \rangle}{\langle \phi^2 \rangle^{2}}$ of the scalar PDFs  generated by the RA forcing method for various target dissipation rates. Pure Gaussian kurtosis of 3.0 and skewness of 0 are denoted by horizontal dashed lines. Larger values of kurtosis are produced for smaller values of $\chit$ while the skewness remains negligible for all $\chit$ values considered. 
Gaussian value of kurtosis is recovered by the method when $\chit \sim 0.01$. In terms of non-dimensional quantities, this value corresponds to a stationary value of $\sim 3.7$ for the scalar and energy turnover time scale ratio defined by

\begin{equation}
r=\frac{2 k\ \chit}{\langle \phi^2 \rangle \epst }
\end{equation}

Figure \ref{fig:moments}(\emph{b}) shows the
 PDF profiles during the stationary period for four values of $\chit$, $0.00046$, $0.01$, $0.129$ and $ 0.932$, corresponding to time scale ratio values of $3.2$, $3.7$, $4.2$, and $8.3$, respectively.  
 As the scalar dissipation rate increases, the scalar field derivatives are larger, indicating the presence of more unmixed regions, with scalar values closer to the bounds. Thus, when $\chit=0.00046$, the scalar PDF is more stretched than a Gaussian, with a kurtosis of $3.7$. $\chit=0.01$ yields a stationary quasi-Gaussian PDF, while $\chit=0.129$ results in a much flatter PDF (quasi-uniform), with $K=1.9$. A further increase of $\chit$ to $0.932$ produces a double delta PDF with peaks near the forcing bounds and $K=1.2$. 
These results correspond to $m=1$ and $n=1$, Taylor Reynolds number $Re_{\lambda} \sim 92$, and $\eta k_{max} =3.0$.    

\subsubsection{Effects of varying stochiometric coefficients}
\label{sec:mn}

\begin{figure}
\begin{center}
(\emph{a}) \hspace{5.9cm} (\emph{b}) \\
\includegraphics[scale=0.3]{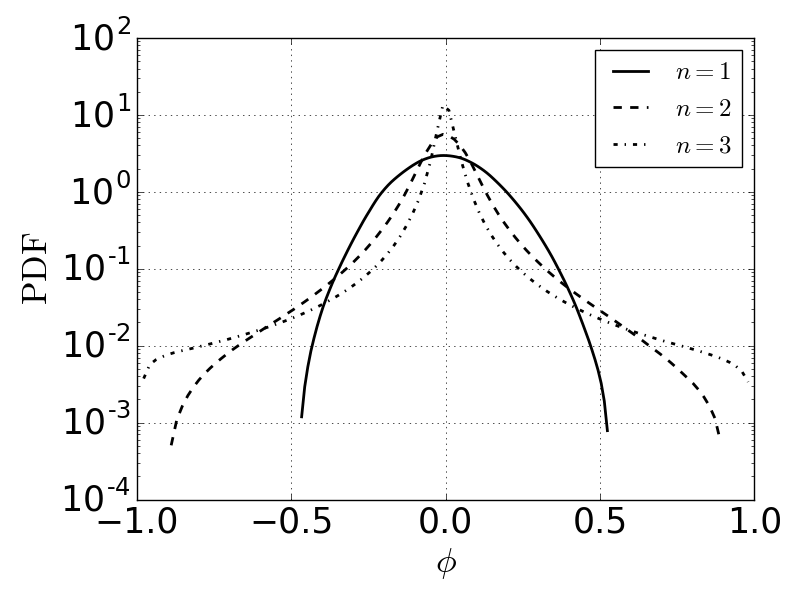}
\includegraphics[scale=0.3]{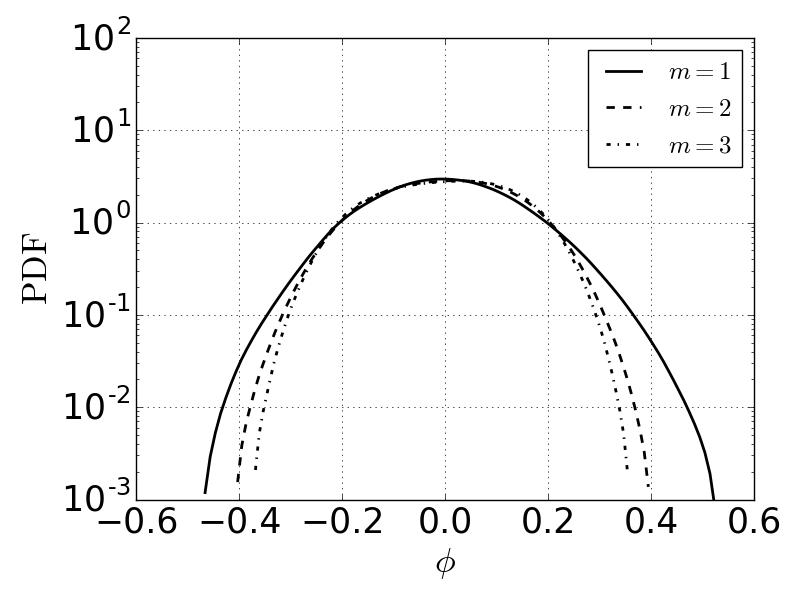}
\end{center}
\caption{Effects of stochiometric coefficients on the scalar PDF at $\chit=0.01$. (\emph{a}) $m=1$, $n=1,\ 2,\ 3$, (\emph{b}) $n=1$, $m=1,\ 2,\ 3$. }
\label{fig:stochio}
\end{figure}

Figure \ref{fig:stochio}(\emph{a}) illustrates the scalar PDFs for the stochiometric
coefficient $m=1$ and  three values of $n$, $n=1,\ 2,\ 3$, for $\epst=0.01$.  
When $n=1$, the resulting scalar PDF is quasi-Gaussian, as explained above. When $n=2$,
the PDF becomes a stretched exponential, with $K\sim 13$. The kurtosis continues to increase and reaches a value of  $\sim 30$ for $n=3$. The results are consistent with the forcing term dependence on the scalar values shown in figure \ref{fig:forcings1}(\emph{c}). Thus, as the coefficient $n$ is increased, the forcing acts more on the richer $A$ or $B$ mixtures and less on the fully mixed fluid, allowing the existence of more partially mixed fluid. This results in more elongated tails in the scalar PDF. On the contrary, when the coefficient $m$ is increased,  figure \ref{fig:forcings1}(\emph{c}) shows that the forcing term is smoother at higher scalar magnitudes and acts stronger near the fully mixed fluid. This removes more of the partially mixed fluid, leaving a larger amount of fluid near the fully mixed value of $\phi=0.0$. 
The result is scalar PDFs with smaller kurtosis values, $2.4$ and $2.3$ for $m=2,\ 3$, respectively (figure \ref{fig:stochio}\emph{b}).

\subsubsection{Effects of choosing spectral resolution and forcing bounds}

\begin{figure}
\begin{center}
(\emph{a}) \hspace{5.9cm} (\emph{b}) \\
\includegraphics[scale=0.3]{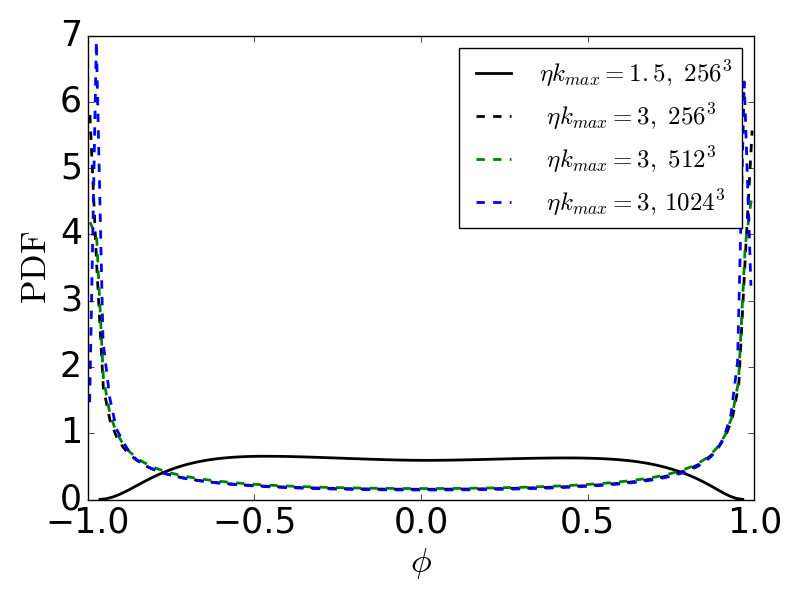}
\includegraphics[scale=0.3]{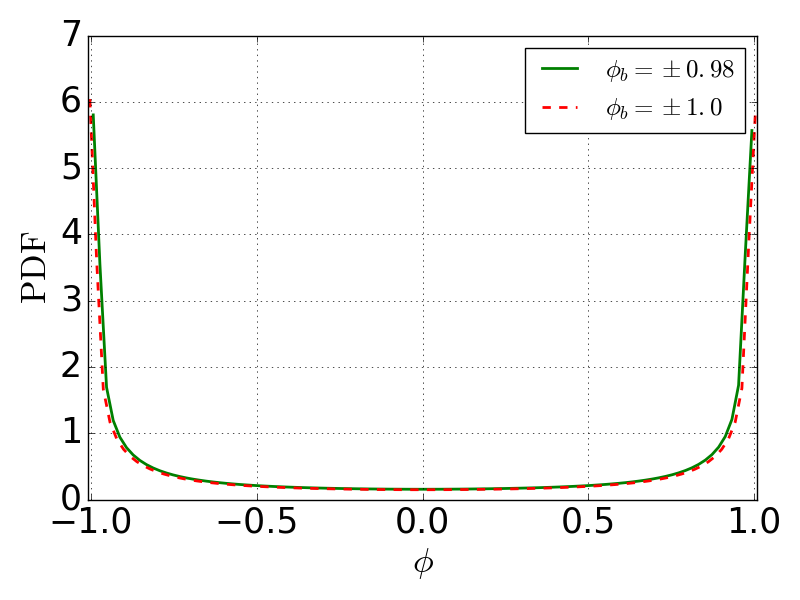}
\end{center}
\caption{(\emph{a}) Scalar PDFs obtained for $\chit$ close to its the largest $\chit$ value supported on a $256^3$ mesh with $\eta k_{max}=3.0$ ($\chit=0.932$) and $\eta k_{max}=1.5$ ($\chit=0.17$). For comparison, scalar PDFs from $512^3$ and $1024^3$ simulations with $\chit=0.932$ are also shown. (\emph{b}) Comparison of double-delta scalar PDFs with different forcing bounds for $\chit=0.932$ on a $256^3$ mesh.}
\label{fig:spectral}
\end{figure}

For a given grid resolution, there is a maximum $\chit$ beyond which the scalar fields are ill-resolved, as the scalar gradients increase with $\chit$. Similarly, for a fixed value of $\chit$, as the resolution
(i.e. $\eta k_{max}$) decreases, at some point the scalar gradients can no longer be represented on the mesh. Figure \ref{fig:spectral}(\emph{a}) illustrates the scalar PDFs corresponding to a scalar dissipation rate close to its maximum allowable value on a $256^3$ mesh, $\chit=0.932$ for $\eta k_{max}=3.0$ ($\Delta x \sim 2 \eta$) and $\chit=0.17$ for $\eta k_{max}=1.5$ ($\Delta x \sim \eta$). The results show that a double-delta scalar PDF can only be obtained if the resolution is large enough to support the relatively large scalar gradients associated with this PDF. As the mesh size is increased, while maintaining   $\eta k_{max}=3.0$ and $\chit=0.932$, the double-delta scalar PDF gets larger peaks. 

In practice, as $\chit$ is increased above a certain threshold (the maximum $\chit$ value referred to above) for a given resolution, scalar values outside their bounds appear in the flow. This threshold can be increased by using forcing bounds magnitudes smaller than the scalar bounds (figure \ref{fig:spectral}\emph{b}). For example, by decreasing the forcing bounds from $\{-1,\ 1\}$ to $\{-0.98,\ 0.98\}$, while maintaining the scalar field bounds at $\{-1,\ 1\}$, the maximum allowable $\chit$ value can be increased. This allows the generation of double-delta PDFs on coarser resolutions. 

\subsection{Comparison with classical methods}

\begin{figure}
\begin{center}
(\emph{a})  \hspace{5cm}         (\emph{b}) \hspace{5cm} (\emph{c}) \\
\includegraphics[width=5cm]{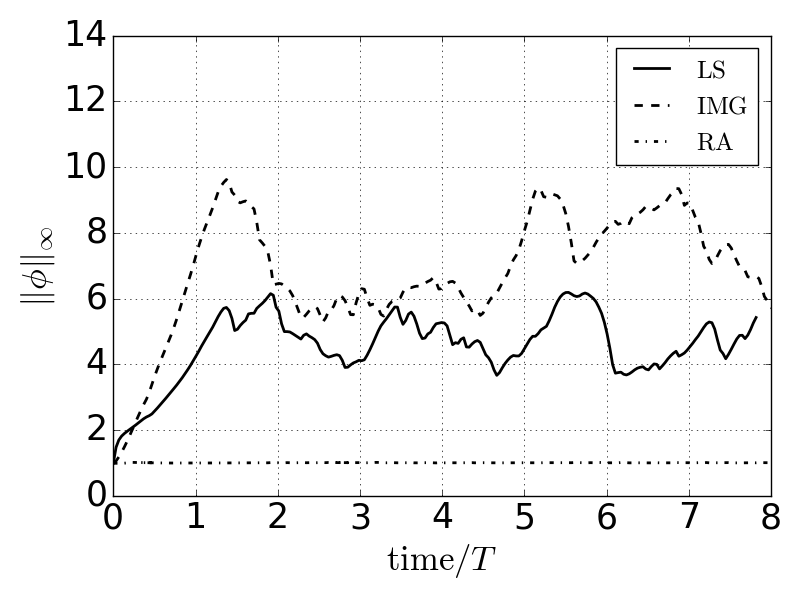}
\includegraphics[width=5cm]{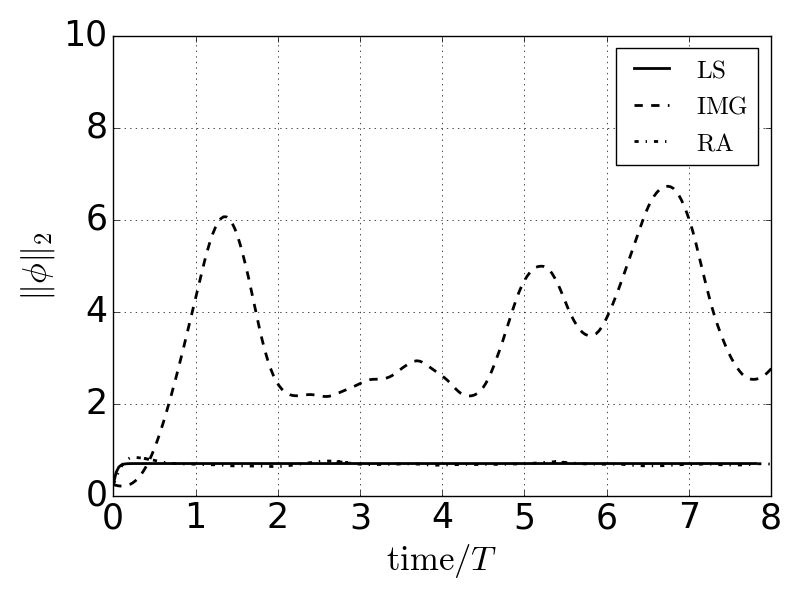}
\includegraphics[width=5cm]{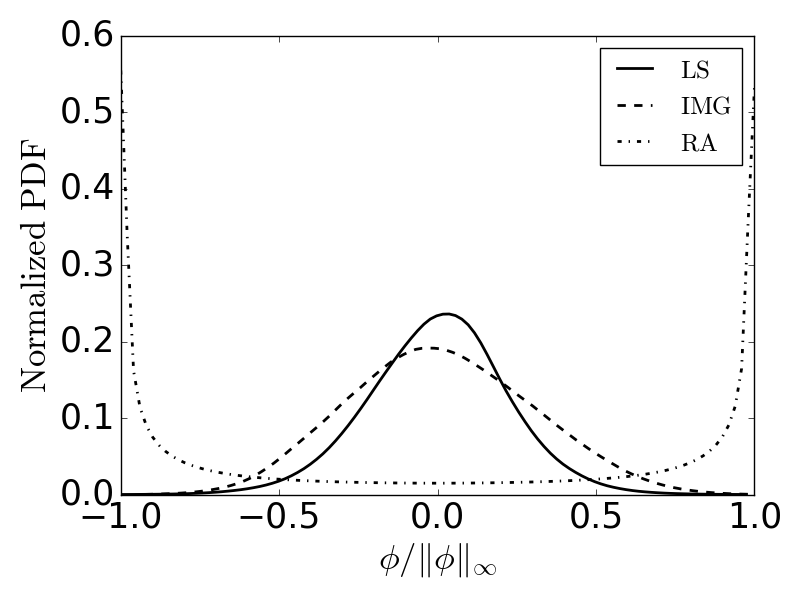}
\end{center}
\caption{(\emph{a}) The $L_\infty$ norm  and (\emph{b}) the $L_2$ norm of of the scalar field using the RA, IMG, and LS methods. The x axis time is scaled with T, the eddy turnover time ($\sim$ 3.0).
(\emph{c}) Stationary scalar PDFs for the three forcing methods, highlighting the ability of the RA forcing to produce non-Gaussian scalar PDFs.}
\label{fig:bounds}
\end{figure}
The ability to produce statistically steady double-delta scalar PDFs
distinguishes the RA method from any previous methods such 
as the imposed mean gradient method, IMG,  of \cite{overholt_pope_1996} or the linear scalar forcing, LS, of \cite{CSB2013}. 
In addition, the RA method has ability to keep the scalar field within prespecified bounds.
Figure \ref{fig:bounds}(\emph{a}) illustrates the temporal evolution of the $L_\infty$ norm 
of the scalar field generated by RA, IMG, and LS methods.
The simulations use identical velocity fields with $Re_\lambda \sim 92$ and $\eta k_{max}=3.0$.
The RA method uses $\chit=0.932$, which gives a double-delta scalar PDF.  
The mean scalar gradient used by the IMG method was chosen to be $1.0$, while the forcing coefficient used by the LS method was chosen to match the scalar variance of the RA scalar.   
As figure \ref{fig:bounds}(\emph{a}) shows, the RA method  keeps the scalar field within the specified bounds, while for the IMG and LS methods, the scalar can take much larger values. For classical methods, the scalar extrema become larger by increasing the values of the forcing parameters such as the scalar variance, the scalar dissipation rate, or the imposed mean scalar gradient. 

The LS forcing method is constructed to attain a specified scalar variance, as shown in figure  \ref{fig:bounds}(\emph{b}). There are no a priori analytical estimates for the scalar variance obtained through the IMG method. Also, figure \ref{fig:bounds}(\emph{c}) highlights the ability of the new RA forcing method to be able to obtain stationary double-delta scalar PDFs, while the RA and IMG methods are restricted to quasi-Gaussian PDFs. By varying the means scalar gradient for the IMG method and forcing strength for the LS method, the scalar PDF can depart slightly from a Gaussian
(not shown), especially for the LS method where only certain parameters lead to a full Gaussian distribution, with a slightly stretched or compressed Gaussian being obtained in the general case.

\subsection{Spectral budget of scalar variance} 

\begin{figure}
\begin{center}
(\emph{a}) \hspace{5.9cm} (\emph{b}) \\
\includegraphics[scale=0.3]{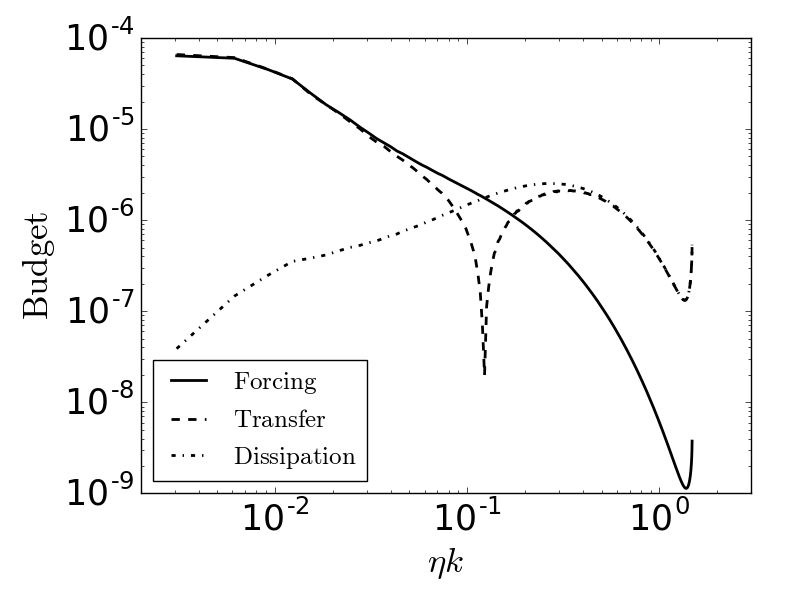}
\includegraphics[scale=0.3]{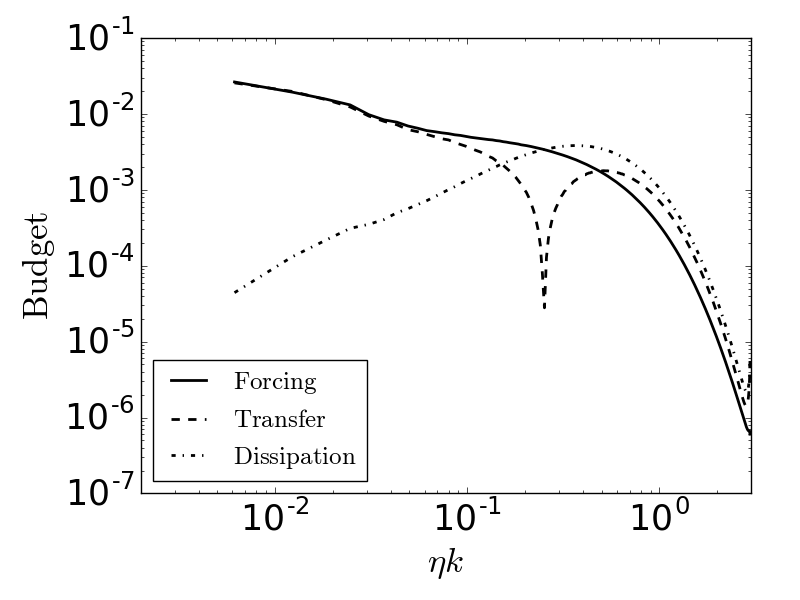}
\caption{Absolute values of the spectra of the terms contributing to scalar variance equation. 
(\emph{a}) $\epst=0.01$, $\eta k_{max}=1.5$, $Re_\lambda=410$, quasi-Gaussian scalar PDF, (\emph{b}) $\epst=0.932$, $\eta k_{max}=3$, $Re_\lambda=255$, double-delta scalar PDF.}
\end{center}
\label{fig:budget}
\end{figure}

In order to illustrate the scale at which the scalar forcing acts, figure \ref{fig:budget} depicts the spectra of the terms in the scalar variance equation:
\begin{equation}
\frac{ \partial{{\phi^2}/{2}}}{\partial t}+ 
\underbrace{\uv . \nabla (\phi^2/2)}_{\mathrm{Transfer}}= 
\underbrace{D \nabla^2 \phi^2/2}  _{\mathrm{Diffusion}}
-  \underbrace{D \nabla \phi \nabla \phi}_{\mathrm{Dissipation}} + 
\underbrace{f_\phi \phi}_{\mathrm{Forcing}}  
\end{equation}
At low $\chit$ values, RA forcing acts mostly at large scales, where it balances the transfer term, so that it leaves the small scalar scales largely unaffected. This results in a quasi-Gaussian or more stretched scalar PDF. However, for large values of $\chit$, in order to keep the scalar near its bounds, forcing needs to locally counteract the smoothing effects of the scalar dissipation and, as figure 
\ref{fig:budget}(\emph{b}) shows, it remains important throughout the dissipation range.

\subsection{Recovering passive scalar dynamics} 

The idea of universality of statistical laws for turbulent fluctuations has been central for fundamental turbulence research for the past several decades. In general, numerical \cite{Watanabe2004,GWS11,gotoh_watanabe_2015} and experimental \cite{Lepore2009} evidence
indicates that the scaling exponents of the higher order scalar structure functions depend on the injection mechanism, i.e. low wavenumber Gaussian white noise temporal forcing for the former and initial conditions for the latter. Thus, scalar scaling satisfying the universal equilibrium theory, if practically realizable, may only be achieved under rather special circumstances. In practical applications, for example in non-premixed combustion, the non-Gaussian statistics of the scalar fields may significantly increase the intermittency and modify even more the scaling behavior.

Here, we explore the ability of the new forcing method to recover the large Reynolds and P\'eclet numbers scaling of the mixed velocity scalar structure function (MSF). Yaglom's equation \cite{Yaglom1949} predicts that MSF should scale as \cite{Watanabe2004,KanedaIshihara2006,GWS11,DAB12}:

\begin{equation}
-\frac{\langle (\Delta_r \phi) ^2 \Delta_r u_r \rangle }{ \bar{\chi } r}=\frac{4}{3},
\end{equation}
in regions much larger than the Batchelor scale and smaller than the energy containing scales. Here,
the scalar and longitudinal velocity differences between two spatial points separated by the displacement $\boldsymbol{r}$, with $r=\|\boldsymbol{r}\|$, are defined by $ \Delta_r \phi= \phi(\xv+ \boldsymbol{r} ) -\phi(\xv), $ and $ \Delta_r u_r = (\uv(\xv+ \boldsymbol{r})-\uv(\xv)).\boldsymbol{e_r}$,  where $\boldsymbol{e_r}$ is a unit vector in the direction of displacement. The MSF results have been verified using both time averaging and a spherical averaging scheme \cite{taylor_kurien_eyink_2003} to extract the isotropic statistics from a single flow snapshot.

\begin{figure}
\begin{center}
(\emph{a}) \hspace{5.9cm} (\emph{b}) \\
\includegraphics[scale=0.3]{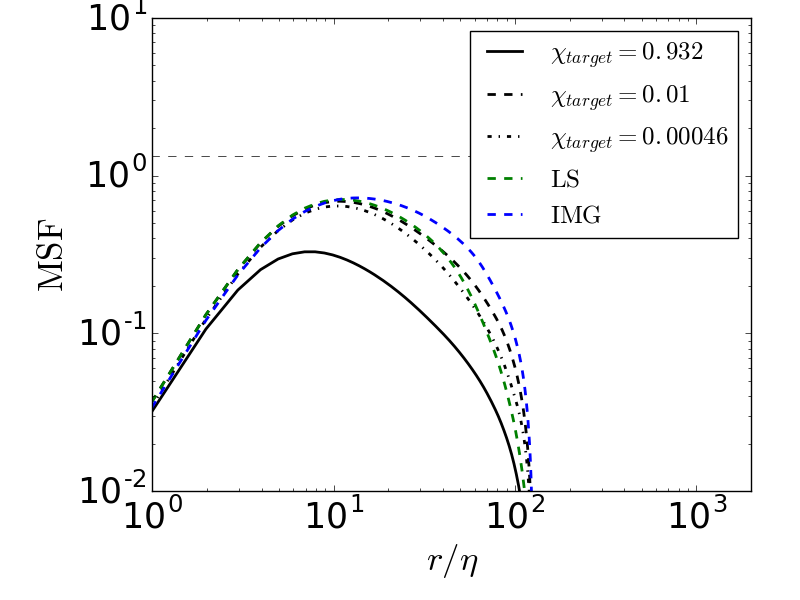}
\includegraphics[scale=0.3]{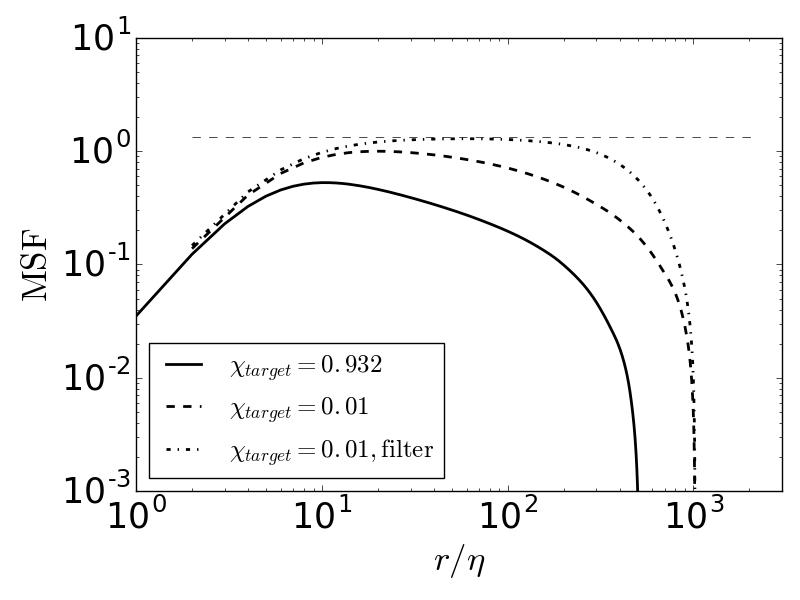}
\end{center}
\caption{Mixed scalar-velocity structure function (MSF) (\emph{a})  $Re_\lambda~92$, with different forcing mechanisms; RA forcing corresponds to double-delta ($\chit=0.932$), quasi-Gaussian 
($\chit=0.01$), and stretched ($\chit=0.00046$) scalar PDF. (\emph{b}) RA forcing for double-delta ($\chit=0.932$), $Re_\lambda=255$, and quasi-Gaussian ($\chit=0.01$), $Re_\lambda=410$, scalar PDFs. The "filtered" results correspond to a low wavenumber restriction of the RA forcing.  }
\label{fig:msf}
\end{figure}

In general, the MSF values are close to LS and IMG forcing results, when a quasi-Gaussian scalar PDF is produced using the RA method (figure \ref{fig:msf}\emph{a}). Simulations with large values of the scalar dissipation rate, with double-delta scalar PDF as the limiting case, consistently yield lower MSF values at all Reynolds numbers examined. Slightly smaller MSF values are also obtained for small scalar dissipation, when the resulting scalar PDF is more stretched, with larger kurtosis values. Thus, a quasi-Gaussian scalar PDF seems to be associated with the maximum MSF values for a given Reynolds number. Nevertheless, for each scalar PDF type, the MSF values increase with the Reynolds number. At $Re_\lambda=255$, the MSF values are still far from $4/3$ for double-delta scalar PDF (figure \ref{fig:msf}\emph{b}). When the RA forcing is restricted to low wavenumbers and the scalar PDF is quasi-Gaussian, Yaglom's law is fully recovered. In this case, the results are consistent with IMG forcing results using similar low wavenumber restrictions and Reynolds numbers  \cite{GWS11}.

\section{Conclusions}
We have presented a novel forcing method for producing stationary scalar fields in incompressible turbulence. The forcing term is constructed based on a hypothetical chemical reaction that transforms ``mixed fluid" (i.e. fluid where the scalar value is closed to its average) back into its ``unmixed components" or ``pure states" (i.e. fluid where the scalar value is closed to its predefined bounds). The reaction form is chosen to ensure that the forcing term satisfies mass conservation, is smooth in the scalar space, and unbiased with respect to the two pure states. By construction, the new forcing term leads to scalar fields that stay within predefined bounds, unlike previous methods that can violate naturally existing bounds, while also generating more general scalar PDFs than previous methods.  

To highlight some of the features of the new forcing method, pseudo-spectral stationary homogeneous isotropic turbulence simulations on $256^3$, $512^3$, and $1024^3$ meshes are presented,  resulting in Taylor Reynolds numbers ranging from $92$ to $410$. The velocity field is forced in the simulations using a low wavenumber restriction of the linear forcing method \cite{Lundgren2003,Rosales_Meneveau05pf,Petersen2010}.  

By varying the target scalar dissipation, which controls the strength of the forcing term, the scalar PDF at stationarity can be changed to cover a large range of kurtosis values, covering stretched exponential, quasi-Gaussian, approximately flat, and double-delta PDF. Additional control  on the shape of the scalar PDF can be exerted through the stoichiometric coefficients of the reactants (mixed and excess pure fluid) in the hypothetical reaction. Thus, the forcing term acts stronger on the less mixed fluid when the coefficient of the excess pure fluid reactant is larger, and stronger in regions where the scalar is near its average value when the stoichiometric coefficient of the mixed fluid reactant is increased. The result is more elongated scalar PDF tails for the former and a narrower scalar PDF for the latter cases. 

As the scalar PDF becomes close to double-delta, the local scalar gradients increase, which is reflected in the larger values of the target scalar dissipation required. In this case, to maintain the separation between fluid states closer to the scalar bounds, the forcing term has a stronger effect at smaller scales. Indeed, the spectra of the forcing term indicate that it is the same order as the scalar dissipation, in the viscous range, when a double delta scalar PDF is maintained. On the contrary, the forcing term is much smaller than the scalar dissipation, in the viscous range, for quasi-Gaussian scalar PDFs. For a given mesh size, double-delta scalar PDFs require more stringent resolution conditions (i.e. large $\eta k_{max}$) and/or tighter scalar bounds in the forcing term 
(i.e. forcing bounds magnitudes smaller than the actual bounds magnitudes) than PDFs with larger kurtosis values. 

For all scalar PDFs considered, the third order mixed scalar-velocity structure function (MSF) increases as the Reynolds number is increased. Quasi-Gaussian scalar PDFs result in MSF values similar to those produced by previous methods, and seem to yield the largest MSF values for a given Reynolds number. Thus, simulations with double-delta scalar PDFs result in markedly lower MSF values, while more stretched scalar PDFs produce slightly lower MSF values than quasi-Gaussian scalar PDFs.  The $Re_\lambda=410$ results with quasi-Gaussian scalar PDF and scalar forcing restricted to small wavenumbers fully recover Yaglom's 4/3 scaling.  Nevertheless, the convergence to Yaglom's scaling is much slower for other scalar PDFs, especially double-delta PDF. 
 
Previous studies indicate that scalar scaling following the universal equilibrium theory may be obtained only under rather special circumstances, which may not be practically realizable. In many practical applications, the scalar PDF may be far from Gaussian (e.g. in non-premixed combustion), which could significantly change the intermittency behavior and scaling of  the scalar fields. While we have not explored such scalings here, we hope that the new tool we have introduced could start shedding some light on the fundamental properties of scalar mixing closer to the practical applications.   

\bibliographystyle{plainnat} 
\bibliography{manuscript_v2}

%\printbibliography

\end{document}